\documentclass[twocolumn,pre,showpacs,floats,superscriptaddress]{revtex4-1}
\usepackage[dvips]{graphicx}
\usepackage{amsmath,bm}
\usepackage{psfrag}
\usepackage{epsfig}
\usepackage{float}

\begin{document}

\title{Stability of a strongly anisotropic thin epitaxial film in a wetting interaction with elastic substrate}

\author{Mikhail Khenner}
\affiliation{Department of Mathematics and Computer Science, Western Kentucky University, Bowling Green, KY 42101}
\author{Wondimu T. Tekalign}
\affiliation{School of Mathematical Sciences, Rochester Institute of Technology, Rochester, NY 14623}
\author{Margo S. Levine}
\affiliation{Department of Radiology at Massachusetts General Hospital and Harvard Medical School, Boston, MA 02114}

\newcommand{\Section}[1]{\setcounter{equation}{0} \section{#1}}
\newcommand{\rf}[1]{(\ref{#1})}
\newcommand{\beq}[1]{ \begin{equation}\label{#1} }
\newcommand{\eeq}{\end{equation} }

\begin{abstract}
The linear dispersion relation for surface perturbations, as derived by Levine \textit{et al.},
%``Self-assembly of quantum dots in a thin epitaxial film wetting an elastic substrate", 
\textit{ Phys. Rev. B}$\;$ {\bf 75}, 205312 (2007) is extended to include a smooth surface energy anisotropy 
function with a variable anisotropy strength (from weak to strong, such that sharp corners and slightly curved facets occur on the corresponding Wulff shape).
Through detailed parametric studies
%The dimensionless perturbation growth rate is thus a function of five key parameters: $h_0$, the thickness of the unperturbed planar film; $k$, the perturbation
%wavenumber; $\mu$, the  ratio of a film to substrate shear modulus;  $\epsilon$, the misfit strain; and $\epsilon_\gamma$, the anisotropy strength.
it is shown that a combination of a wetting interaction and strong anisotropy, and even a wetting interaction alone results in complicated linear stability 
characteristics of strained and unstrained solid films. 
%(Khenner \cite{mypapers}).  
\end{abstract}

\date{\today}
\maketitle

\Section{Introduction}

Studies of the morphological instabilities of a thin solid films are a first step towards understanding complex phenomena such as the formation of a three-dimensional nanoscale islands in 
strained alloy heteroepitaxy.
Such studies became common after the pioneering works of Asaro and Tiller \cite{AT}, Grinfeld \cite{G} and Srolovitz \cite{S}. The classical Asaro-Tiller-Grinfeld instability
is one of an uniaxially stressed solid film on a rigid infinite substrate. Its variants for the single-component and alloy films on the rigid as well as on the deformable substrates have been studied
and this research continues. Reviews of works on the single-component films have been published, see for instance Ref. \cite{GN_SB}.

%Film-substrate wetting interaction is a relatively new concept in the field of research on morphological instability and evolution \cite{CG,SZ,ORS}. 
%%and therefore the theoretical and modeling 
%%exploration of wetting effects on morphological stability of stressed films is only beginning, see for instance \cite{LevinePRB2007,SVD,GillWang,PangHuang}. 
%Since such interaction provides one 
%possible mechanism capable of terminating unlimited pattern coarsening \cite{LevinePRB2007,PangHuang,KE}, the focus in the majority of 
%modeling papers published to-date is, quite understandably, on the analysis of weakly nonlinear instability that gives rise to the formation of a pattern of structures, or on 
%the direct
%numerical simulation of a fully nonlinear problem.

Film-substrate wetting interaction
%, characterized by an additional wetting energy that depends on the local film thickness $h$, 
is a relatively new concept in the field of research on morphological instability and evolution. When surface slopes are not very large, this additional wetting energy can be considered a function of the local film thickness $h$, but not the slopes of $h$ \cite{CG,SZ,ORS}. 
In Refs. \cite{GolovinPRB2004,LevinePRB2007,PangHuang,KE} and others it has been shown that wetting interaction damps long-wave instability modes
in a certain range of film thickness, thus changing the instability spectrum from long-wave type to short-wave type. The latter mode of instability is more relevant to the process of formation 
of island arrays \cite{Chiu}. 
In Ref. \cite{LevinePRB2007} it is recognized that 
in the presence of wetting
interaction, the boundary conditions that describe stress balance at the film free surface and at the film-substrate interface must be augmented by \textit{wetting stress} terms - that are proportional to the rate of change of the surface energy with $h$. (Wetting stress is called conjoining (or disjoining) pressure when studying dynamics of thin \emph{liquid} films on substrates \cite{Book1,Book2}. This pressure is partially responsible for so-called spinodal instability, which typically leads to film dewetting (rupture); for discussions of spinodal instability, see for instance Ref. \cite{Sharma} and references therein.) Wetting stress and lattice-mismatch stress have different origin, and the former may be present even when the latter is absent.
For wetting (non-wetting) films, the solution of the elastic
free-boundary problem with boundary conditions that include wetting stress terms,
results in additional destabilizing (stabilizing) contributions in the dispersion
relation.
Some stability characteristics have been analyzed in Ref. \cite{LevinePRB2007} within the framework of longwave approximation, where in addition the surface energy is assumed isotropic.
This communication extends that work by adding strong anisotropy and considering not only wetting films, but also non-wetting films. 
Here we recognize that the film thickness and
the wetting length are two \emph{independent} characteristic lengths, i.e. the former length is determined by film deposition, while the latter one is determined by the molecular structure and properties of the film-substrate interface. Since wetting length may be, and normally is, less than the deposited film thickness, the perturbation wavelengths may be comparable to the film thickness but still much larger than the wetting length. In this case the long wavelength approximation may hold with respect to the wetting length, but not with respects to the film thickness. 
In Sec. IV we show that this approach reveals linear stability features that, we believe, went unnoticed in prior publications.
%For this reason, where appropriate, the perturbation wavelengths are considered not necessarily large with respect to film thickness.
%In some cases, this approach reveals linear stability features that, we believe, went unnoticed in prior publications.

%We show that very nontrivial linear stability 
%emerges when ... ({\bf to be filled in})
%the film thickness is less than roughly ten times the characteristic wetting length. 

%We remark here that at least one model has been published where wetting stress is not accounted for and thus contributions from wetting potential and strain are decoupled in the dispersion relation 
%\cite{WS,KE}.
%Due to such decoupling the resulting stability characteristics are simpler than those revealed in Ref.  \cite{LevinePRB2007} and in this paper, despite that the substrate is taken deformable and the 
%solution of the elasticity problem is more complicated. 

\Section{Problem statement}

\label{Sec2}

Following Refs. \cite{SpencerJAP93,LevinePRB2007}, we consider a dislocation-free,  one-dimensional, single-crystal, epitaxially strained thin solid film in a wetting interaction with a
solid, semi-infinite elastic substrate. The film surface $z=h(x,t)$ evolves due to surface diffusion. This evolution is described by
\begin{equation}
\frac{\partial h}{\partial t} = \frac{\Omega D N}{kT}\frac{\partial}{\partial x}\left[\left(1+\left(\frac{\partial h}{\partial x}\right)^2\right)^{-1/2}\frac{\partial \mathcal{M}}{\partial x}\right],
\label{1.1}
\end{equation}
where $\Omega$ is the atomic volume, 
$D$ is the adatoms diffusivity, $N$ is the adatoms surface density, $kT$ is the Boltzmann factor,
and $\mathcal{M}$ is the surface chemical potential \cite{MULLINS5759}.
The latter has contributions from the elastic energy in the film, the anisotropic surface energy, and a wetting interaction:
\begin{eqnarray}
\nonumber \mathcal{M} = \mathcal{E}(h)+ 
\Omega \left[\left(\gamma+\frac{\partial^2\gamma}{\partial \theta^2}\right)\kappa-\delta\left(\frac{\kappa^3}{2}+\frac{\partial^2\kappa}{\partial S^2}\right)\right.+\\
\left.\left(\frac{\partial \gamma}{\partial h} - 
{\mbox sign}(h_x)\frac{\partial^2 \gamma}{\partial h \partial \theta}\right)\cos{\theta}\right],
\label{1.4e}
\end{eqnarray}
where 
$\theta$ is the angle that the unit surface normal makes with the reference crystalline direction, say [01]
(chosen along the $z$-axis, which is normal to the substrate),
$\gamma(h,\theta)$ is the height- and orientation-dependent surface energy,
%of a wetting layer, 
$\kappa$ is the curvature of the surface, $S$ is the arclength along the surface and $h_x$ is surface slope (note, $\partial/\partial S = (\cos{\theta}) \partial/\partial x = 
\left(1+h_x^2\right)^{-1/2}\partial/\partial x$).
%Angle $\theta$ measures the orientation of the surface with respect to the underlying crystal structure.
The term proportional to the small positive parameter $\delta$ is the regularization that is required in view of ill-posedness of Eq. \rf{1.1} for strong anisotropy, that is when 
$\epsilon_\gamma \ge 1/(m^2-1)$ in Eq. \rf{1.2} below \cite{HERRING1,AG,CGP,Spencer}. Note also that the mixed derivative term in Eq. \rf{1.4e} is nonlinear and thus it has no impact on linear
stability. 

In this communication we consider the two-layer exponential model for the surface energy \cite{CG,ORS}:
%of a wetting layer reads \cite{CG}:
\begin{equation}
\gamma(h,\theta) = \gamma_t(\theta) + \left(\gamma_s-\gamma_t(\theta)\right)\exp{\left(-h/\ell\right)},
\label{1.4fp}
\end{equation}
where $\gamma_s$ is the surface energy of the substrate when there is no film,  $\ell$ is characteristic wetting length, and $\gamma_t(\theta)$ is the anisotropic surface energy of a thick film:
\begin{equation}
\gamma_t(\theta) = \gamma_0 (1+\epsilon_\gamma \cos{m\theta}),\quad \epsilon_\gamma \ge 0.
\label{1.2}
\end{equation}
Here $\gamma_0$ is the mean surface energy, $\epsilon_\gamma$ is the strength of anisotropy and $m$ is the integer parameter specifying anisotropy type (i.e., four-fold, six-fold, etc.) By comparison with experimental and \textit{ab initio} computational 
studies the two-layer exponential model has been shown the most accurate 
to-date \cite{BWA,GillWang}.  In the absence of anisotropy, $\gamma_t=\gamma_0=const.$, and $\delta$ vanishes. $\delta$ is taken zero also at weak anisotropy, $\epsilon_\gamma < 1/(m^2-1)$.
In reality, the maximum of $\gamma_t(\theta)$ might occur at $\theta=\beta$, where the non-zero angle $\beta$ is a misorientation from the reference direction.
Without significant loss of generality we assume $\beta=0$ in Eq. \rf{1.2}.

The expression for the elastic energy $\mathcal{E}(h)$ in Eq. \rf{1.4e} is derived in Ref. \cite{SpencerJAP93} without accounting for wetting interaction. 
Wetting interaction is considered in several papers, including
Refs. \cite{WS,GolovinPRB2004,LevinePRB2007,PangHuang,KE,GillWang,mypapers,EK} (in Ref. \cite{EK} the wetting effect arises not from the dependence of surface energy on thickness, 
but from the thickness-dependent elastic energy, which cannot be calculated from linear elasticity theory). 
To this end, by combining expressions derived by us in Refs. \cite{LevinePRB2007,mypapers} we state the dimensionless
linear growth rate \textit{in the longwave limit}, $kh_0\ll 1$ (where $k$ is the perturbation wavenumber and $h_0$ is the uniform thickness of unperturbed planar film):
\begin{eqnarray}
\nonumber \omega(h_0,k,\mu,\epsilon,\epsilon_\gamma)=A\epsilon^2\left(\mu+A_1 h_0 k\right)k^3-\\
\nonumber B \epsilon \left[\mu\left(h_0-1\right)+h_0\left(B_1 h_0-A_1\right) k\right]k^3 a \exp{\left(-h_0\right)}+\\
\nonumber F\left[\left(\Lambda-(G+\Lambda)\exp{\left(-h_0\right)}\right)k^4-\Delta k^6-\right.\\
\left.ak^2\left\{\exp{\left(-h_0\right)} - B_2a\exp{\left(-2h_0\right)}\right\}\right].
\label{omega_low}
\end{eqnarray}
$\ell$ has been chosen as the length scale and $\ell^2/D$ as the time scale. 
%(since we are concerned with linear stability in this paper, such simple choice of scales
%facilitates analyses for varying film thickness with respect to wetting length).  
Here 
%$k$ is the wavenumber of the perturbation, 
$\epsilon$ is the misfit strain in the film, and $\mu=\mu_f/\mu_s$ is the ratio of the film shear modulus to the substrate shear modulus. 
%and $h_0$ is the uniform thickness of unperturbed planar film. 
Other parameters are:
\begin{equation}
A = \frac{8 N \Omega^2 \alpha_s \left(1+\nu_f\right)^2 \mu_f}{kT \ell \alpha_f^2},\quad B =  \frac{4 N \Omega^2 \gamma_0 \alpha_s \left(1+\nu_f\right) \nu_f}{kT \ell^2 \alpha_f^2},
\label{Z_X}
\end{equation}
\begin{equation}
A_1 = \frac{2C_1}{\alpha_f \alpha_s},\quad B_1 = \frac{\nu_fC_2-\alpha_f}{2\alpha_f\alpha_s \nu_f},\quad B_2 = \frac{\beta_f\gamma_0}{\alpha_f\mu_f\ell},
\label{Z1_X1}
\end{equation}
\begin{equation}
C_1 = \alpha_f+\alpha_f \beta_s \mu-\alpha_s^2 \mu^2,\quad C_2 = 4\alpha_f+3\alpha_f \beta_s \mu-4 \alpha_s^2 \mu^2,
\label{C1_C2}
\end{equation}
\begin{equation}
\alpha_{f(s)} = 2\left(1-\nu_{f(s)}\right),\quad \beta_{f(s)} = 1-2\nu_{f(s)},
\label{alpha_beta}
\end{equation}
\begin{equation}
F = \frac{N\Omega^2 \gamma_0}{kT \ell^2},\quad 
\Delta = \frac{\delta}{\gamma_0 \ell^2},
\label{B_D}
\end{equation}
\begin{equation}
\Lambda = \left(m^2-1\right)\epsilon_\gamma -1,\quad G=\gamma_s/\gamma_0,\quad a= G-1-\epsilon_\gamma.
\label{L_G_a}
\end{equation}
In Eqs. \rf{Z_X}-\rf{alpha_beta} $\nu$ is Poisson's ratio. 
Note coupling of wetting interaction and misfit strain through the term proportional to $\epsilon$ (the second line of 
Eq. \rf{omega_low}). This term, responsible for breaking symmetry between compressive and tensile stress states, drops out of the growth rate in the absence of wetting interactions, $h_0\rightarrow \infty$. In the square brackets of this term, $-\mu$ and $-A_1h_0k$ are the contributions from the wetting stress; see also Refs. \cite{LevinePRB2007,mypapers}. Another contribution from the 
wetting stress is the term proportional to $\exp{(-2h_0)}$ in the square brackets in the last line of the equation. 
%If wetting stress is omitted from the formulation of the elasticity problem, then 
%the only terms due to wetting interaction that remain in the equation are $\epsilon B\mu h_0k^3a\exp{\left(-h_0\right)}$, $\epsilon BB_1h_0^2k^4a\exp{\left(-h_0\right)}$ 
%and $-Fak^2\exp{\left(-h_0\right)}$ (see \cite{LevinePRB2007,mypapers} for more information).  
%This scale as well as the time scale differ from the conventional scales \cite{SpencerJAP93}.
%Some contributions are  omitted from Eq. \rf{omega_low}.
%They are the wetting contributions (proportional to $\exp{\left(-h_0\right)})$ and the stress contributions to the regularization ($k^6$-term)
%\cite{mypapers,GolovinPRB2004}. We assume that the omitted contributions to regularization
%are lumped in the adjustible parameter $\delta$. 

Our goal is to elucidate the roles of anisotropy, wetting interaction and
wetting stress and to characterize film stability in the space of dimensionless parameters $h_0$, $k$, $\mu$, $\epsilon$ and $\epsilon_\gamma$. Other material parameters will
be fixed to their most characteristic values. We choose the following values: $D = 1.5\times10^{-6}$ cm$^2$/s, $N= 10^{15}$ cm$^{-2}$, $\Omega= 2\times10^{-23}$ cm$^3$, $kT = 1.12 \times 10^{-13}$ erg,
$\gamma_0 = 2\times10^3$ erg/cm$^2$, $\nu_f = 0.198$, $\nu_s = 0.217$, $\mu_f = 10^{12}$ erg/cm$^3$, $\delta = 5\times10^{-12}$ erg, and $\ell = 3\times10^{-8}$ cm.
The value of the characteristic wetting length is of the order of 1 ML thickness for InAs or Ge film \cite{GillWang}. We assume strong anisotropy, i.e. $\epsilon_\gamma > 1/ \left(m^2-1\right)\equiv \epsilon_\gamma^{(c)}$ and
thus $\Lambda > 0$. For strained films considered in Sec. IV, we choose $m=32$ as the most characteristic value \cite{TSRvK}. However, as far as the effect of anisotropy on linear stability is of interest, 
similar results are obtained  for other common values such as $m=4$ or $m=6$. That is, choosing larger $m$ simply means that smaller values of $\epsilon_\gamma$ are above the critical value 
$\epsilon_\gamma^{(c)}$. \emph{Wetting films} require $a>0$ \cite{mypapers}, thus we choose $\gamma_s = 2\gamma_0,\ \epsilon_\gamma^{(c)}< \epsilon_\gamma < 1$.
For analysis of \emph{non-wetting films} ($a<0$), we choose $\gamma_s = \gamma_0/2,\ \epsilon_\gamma > \epsilon_\gamma^{(c)}$. It is clear that wetting stress terms (pointed out above) are destabilizing
(stabilizing) in wetting (non-wetting) films.

\Section{Films with wetting interaction and zero misfit strain and wetting stress}

\begin{figure}[h]
%\centering
\includegraphics[width=0.8\linewidth]{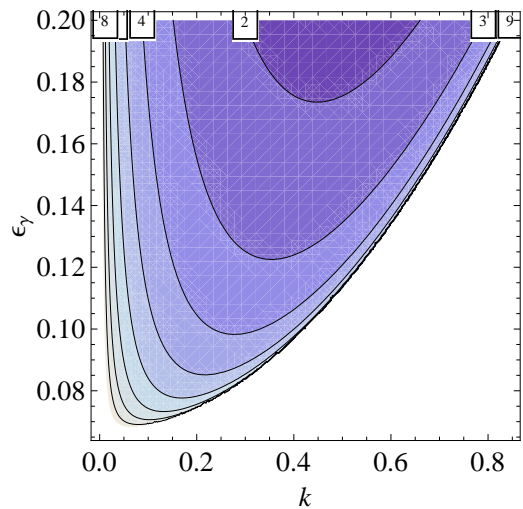}
\caption{Contour plot of the critical thickness $h_0^{(c_1)}$ for strong anisotropy, $\epsilon_\gamma > \epsilon_\gamma^{(c)} = 1/15$.}
\label{Cploth0_c}
\end{figure}

When misfit strain and wetting stress are not present, Eq. \rf{omega_low} reduces to:
\begin{eqnarray}
\nonumber \omega(h_0,k,\epsilon_\gamma)=
F\left[\left(\Lambda-(G+\Lambda)\exp{\left(-h_0\right)}\right)k^4\right.\\
\left.-\Delta k^6-ak^2\exp{\left(-h_0\right)}\right].
\label{omega_low00}
\end{eqnarray}
First, we consider wetting films. 
It follows that the perturbations with the wavenumbers larger than $k_c=\sqrt{\Lambda/\Delta}$ cannot destabilize a film of any thickness. 
%(that is, $\omega<0$ for $k>k_c$ and any $h_0$).
(Here, $k_c$ is not the customary cut-off wavenumber, but is determined from the condition $\omega<0$ for any $h_0$.)
However, in the opposite case $k<k_c$ only the films of thickness that is less than the critical, $h_0^{(c_1)}$, are stable:
\begin{equation}
h_0 < h_0^{(c_1)} = - ln\frac{\Lambda k^2 - \Delta k^4}{a+(G+\Lambda)k^2}.
\label{h0_c1_nostress}
\end{equation} 
With $\Delta = 25/9$ corresponding to the material parameters stated above, $m=4$, and $\epsilon_\gamma = 0.1$, we obtain $k_c = 0.42$. Taking typical $k = 0.1k_c$ in 
Eq. \rf{h0_c1_nostress} gives $h_0^{(c_1)}=6.94$, which translates to the dimensional value of 7 ML. 
Fig. \ref{Cploth0_c} shows the contour plot of $h_0^{(c_1)}(k,\epsilon_\gamma)$. It can be seen that
stronger anisotropy decreases $h_0^{(c_1)}$. 
We notice also that strong anisotropy destabilizes
(that is, the contribution proportional to $k^4$ in the square bracket is positive) only relatively thick films, such that
\begin{equation}
h_0 > h_0^{(c_2)} = - ln\frac{\Lambda}{G+\Lambda}.
\label{h0_c2_nostress}
\end{equation} 
For the chosen values, $h_0^{(c_2)}=1.6$ ML. Such threshold-type influence of strong anisotropy is distinctly different from the simplified model in which wetting
interaction is absent. The latter model can be obtained by taking $h_0\rightarrow \infty$ in 
Eq. \rf{omega_low00}, and thus this equation becomes $\omega(k,\epsilon_\gamma)=F\left(\Lambda k^4-\Delta k^6\right)$, from which it is clear that
strong anisotropy has destabilizing influence on a film of arbitrary thickness.
These findings to some extent echo Refs. \cite{EK,GillWang}, where the 
existence of the critical perturbation amplitude that is necessary to destabilize a film in the presence of a cusp in the surface energy $\gamma(\theta)$
(which is the case below the roughening temperature), has been demonstrated. Thus if a film is thin, critical amplitude may be unattainable and the film will not  be destabilized.
However note that models of Refs. \cite{EK,GillWang} do not allow staightforward separation of the effects of surface energy and mismatch stress, and thus our results
can't be easily compared to these papers.
%However, the effects of surface energy and stress are difficult to separate in Refs. \cite{EK,GillWang} 
%However, it must be noted that the analysis in Refs. \cite{EK,GillWang} includes stress from the outset, and therefore it is difficult to separate the effects of stress and energy 
%there. 

Next, we consider non-wetting films. One example of such material system may be the energetically-driven 
dewetting of silicon-on-insulator \cite{SECS,DSMK}. Repeating the analysis and referring to the critical values shown above, it follows that film of any
thickness is stable with respect to perturbations with wavenumbers larger than $max\left(k_c,k_c^{(u)}\right)$, where $k_c^{(u)}=\sqrt{-a/(G+\Lambda)}$.
If $k_c<k<k_c^{(u)}$, then film is stable if $h_0> h_0^{(c_1)}$ and unstable otherwise. If  $k_c^{(u)}<k<k_c$, then the film is stable if $h_0< h_0^{(c_1)}$ and unstable otherwise.
Finally, if $k<min\left(k_c,k_c^{(u)}\right)$, then the film of any thickness is unstable. With $G=0.5$ and $\epsilon_\gamma=0.1$, $k_c^{(u)}=0.77>k_c$,
and therefore the second possibility, $k_c^{(u)}<k<k_c$, must be dismissed. Typically, the first scenario ($k_c<k<k_c^{(u)}$) holds, and thus 
there is a critical thickness below which the film is unstable \cite{PangHuang}. 
%We will not further consider non-wetting films in this paper.

Results similar to shown above for wetting and non-wetting films can be obtained (numerically) with non-zero wetting stress, since the negative exponent $\exp{(-2h_0)}$ decays fast compared to 
the terms in Eq. \rf{omega_low00} that are proportional to $\exp{(-h_0)}$. 
%For example, Fig. \ref{r-u-eps_gamma}
%
%\begin{figure}[H]
%\centering
%\includegraphics[width=0.9\linewidth]{sec5.eps}
%\caption{Effect of anisotropy on stability for (a)
% $\epsilon_{\gamma}=0.01$, (b) $\epsilon_{\gamma}=0.1$}
%\label{r-u-eps_gamma}
%\end{figure}
%

\Section{Wetting films with non-zero misfit strain and wetting stress}

The situation presented in this section is common for Stranski-Krastanov growth
of epitaxial thin films. 
%We merely attempt here to augment the linear stability results of Ref. \cite{LevinePRB2007} by anisotropic effects 
%and further elucidate the role of wetting interaction. 

%Since the growth rate is quadratic in $\epsilon$, one can explicitly determine the boundaries of neutral stability, $\omega=0$.
%This is done in Ref. \cite{LevinePRB2007}, where such boundaries are plotted in the $h_0-\epsilon$ plane using the longwave growth rate and isotropic surface energy.
%However, the plots presented are for a single perturbation wavenumber, corresponding to the fastest growing mode. 

\begin{figure}[h]
%\centering
\includegraphics[width=0.8\linewidth]{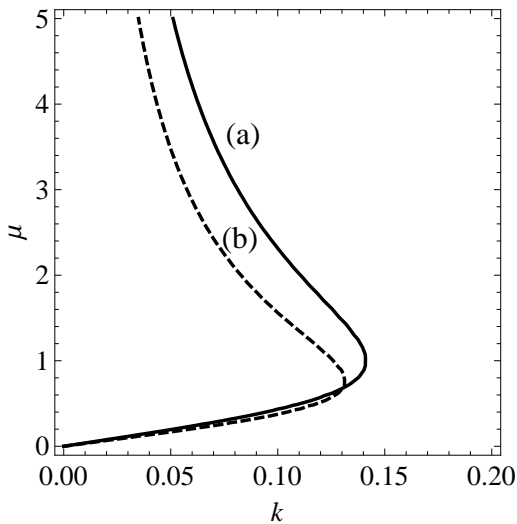}
\caption{Zero level curve of $f(k,\mu,h_0)=\mu\left(h_0-1\right)+h_0\left(B_1 h_0-A_1\right) k$.
(a) $h_0=2$, (b) $h_0=3$. 
%Domains are marked in the $k\mu$-plane where 
To the left of each curve,
the symmetry-breaking term in the longwave growth rate \rf{omega_low} (second line) is stabilizing, to the right - destabilizing 
(when $a,\epsilon >0$).}
\label{coeff1}
\end{figure}

As we pointed out in Sec. II, 
in the presence of misfit strain and wetting interaction, Eq. \rf{omega_low} contains the term that is proportional to the 
first power of misfit strain. Whether this term is destabilizing or stabilizing (for, say, $\epsilon>0$) depends on the sign of the expression
$f(k,\mu,h_0)=\mu\left(h_0-1\right)+h_0\left(B_1 h_0-A_1\right) k$. 
Only for sufficiently small $k$ and large $\mu$ 
%(for instance, $\mu>0.99$ for $h_0=2$ and $k=0.14$) 
this is positive. Then the
second term in Eq. \rf{omega_low} is stabilizing as shown in
Fig. \ref{coeff1}. Increasing $h_0$ makes the domain of stabilization smaller. As $\mu$ is in the range 0.5 - 1.0 for a typical heteroepitaxial semiconductor system, the 
coupling of misfit strain and wetting interaction has stabilizing effect on an ultrathin film of thickness of the order of several wetting lengths, for \emph{longwave perturbations}. Note that the standard $\epsilon^2$-term
is always destabilizing for all perturbation wavelengths \cite{SpencerJAP93}.

In order to demonstrate some effects of arbitrary relation between wetting length, film thickness and the perturbation wavelength, in conjunction with strong anisotropy,
%arbitrary ratio of the perturbation wavelength to the film thickness, 
we use next the full \textit{dimensional} growth rate expression
involving hyperbolic functions of $kh$, where $k,\ h$ are now the dimensional wavenumber and mean thickness, respectively. (Available on request from authors.)  
(Eq. \rf{omega_low} emerges upon expansion of this growth rate in powers of small
dimensionless parameter $kh$, retaining the dominant terms of the expansion (longwave approximation), and non-dimensionalization.) 
As is Eq. \ref{omega_low}, the full growth rate is quadratic in $\epsilon$, allowing one to explicitly determine the boundaries of neutral stability, $\omega=0$, in the $r-\epsilon$ or 
$u-\epsilon$ planes. Here $r$ and $u$ (dimensionless) are defined by $h=r\ell,\ k=u(2\pi/\ell)$. 
%It will be demonstrated below that
%even without anisotropy the stability is more complicated than follows from the analysis presented in Ref. \cite{LevinePRB2007}. 
%Our primary objective is to investigate how the strong surface energy anisotropy influences stability. 

\begin{figure}[h]
%\centering
\begin{tabular}{cc}
\epsfig{file=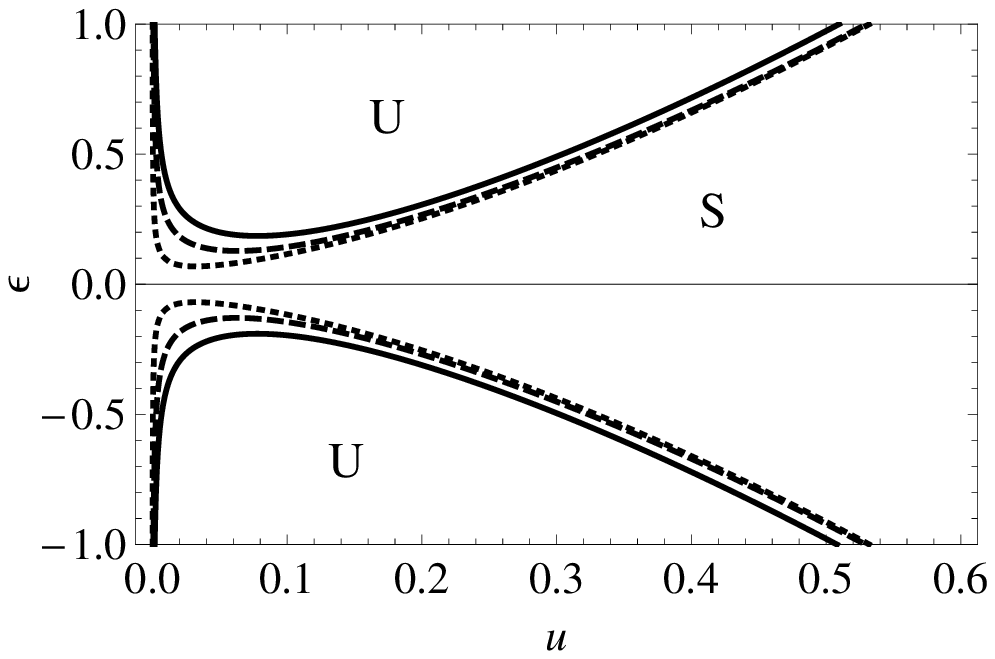,width=0.45\linewidth} &
\epsfig{file=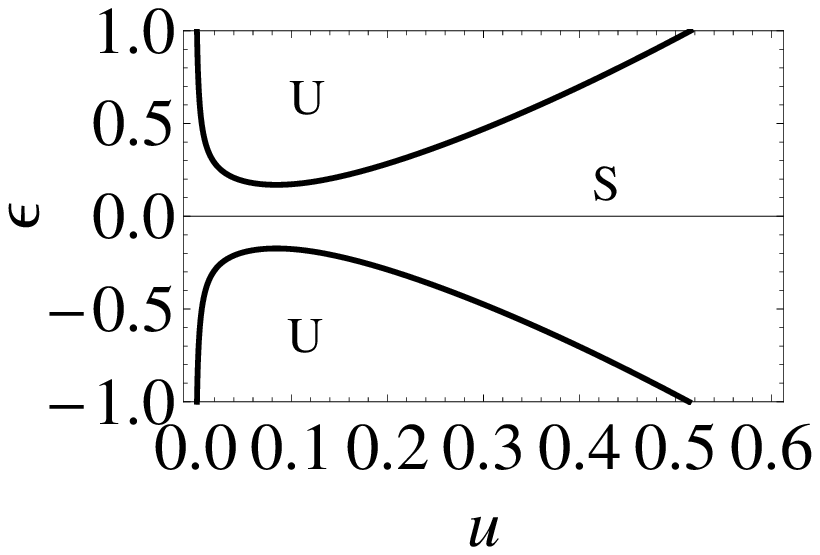,width=0.45\linewidth} \\ [2ex]
(a) & (b)\\ [2ex]
\epsfig{file=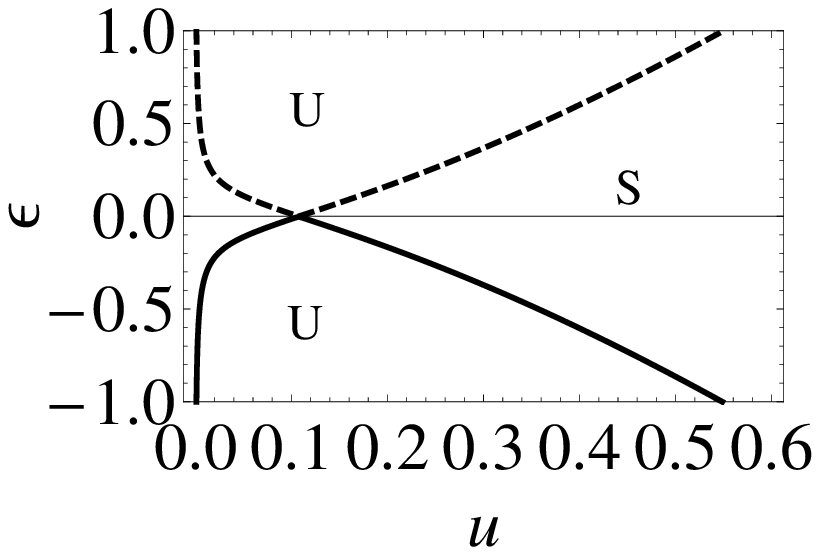,width=0.45\linewidth} & 
\epsfig{file=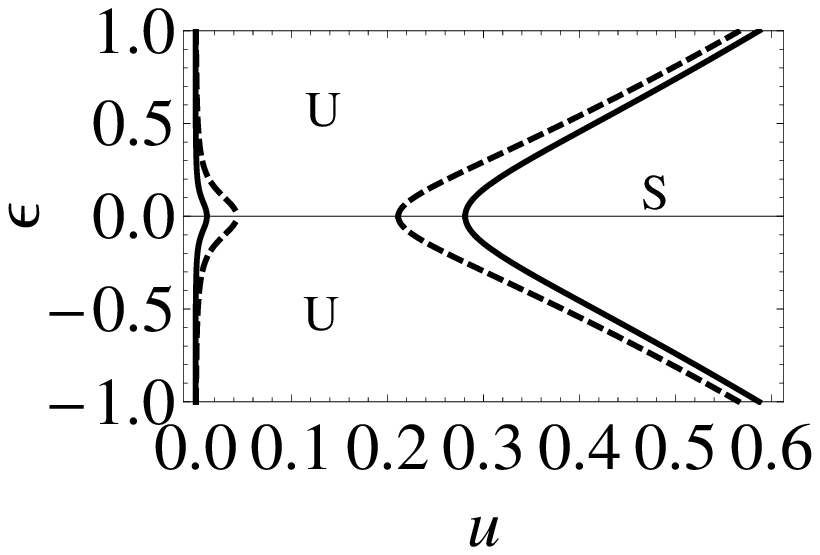,width=0.45\linewidth} \\ [2ex]
(c) & (d) \\ [2ex]
\end{tabular}
\caption{Neutral stability curves. (a): $\epsilon_\gamma=0$. $r=0.1$ (solid), $r=1$ (dash), $r=3$ (dot).  
(b-d): $\epsilon_\gamma=0.01$. (b) $r=0.1$, (c) $r=0.518$, (d) $r=1$ (dashed), $r=3$ (solid). Domains of surface stability (instability) are
marked by S (U).
%Film is stable in the domains bounded by two solid curves symmetric about the $u$-axis.
}
\label{epsg0_01}
\end{figure}

\begin{figure}[h]
%\centering
\includegraphics[width=0.8\linewidth]{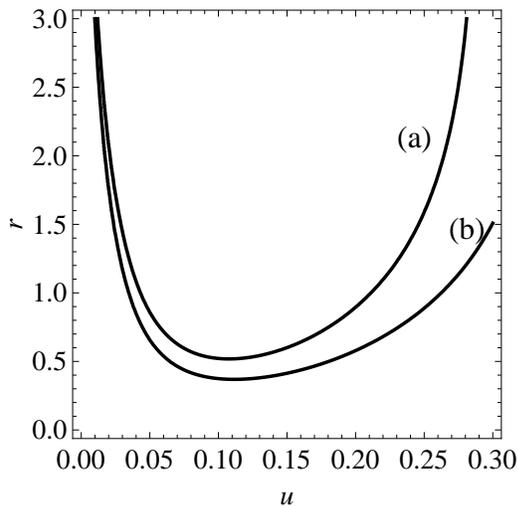}
\caption{Neutral stability curves (see text). (a)
 $\epsilon_{\gamma}=0.01$, (b) $\epsilon_{\gamma}=0.014$. Film is stable below each curve.}
\label{spacing}
\end{figure}

Fig. \ref{epsg0_01} shows neutral stability curves, in $u-\epsilon$ plane, for $\epsilon_\gamma =0$ and 0.01. (For value $m=32$ used in this Section, $\epsilon_\gamma^{(c)} = 
0.001$.)
For all three values of a film thickness in the former (isotropic) case, and for the smallest value in the latter (strongly anisotropic) case, 
the film is destabilized by \emph{short-wavelength} perturbations, $u>0.02$, above some critical value of the misfit parameter $\epsilon$.
Increasing film thickness in the isotropic case to values as large as $50\ell$ only makes the domain of stability to shrink.
However, for larger film thickness in the strongly anisotropic case (Fig. \ref{epsg0_01}(c,d)) 
two stability domains emerge separated by the domain of instability. The splitting of a single domain into two domains occurs at $r=0.5$. 
The size of stability domains decreases with increasing film thickness.
Overall, the film is less stable with increasing anisotropy (as expected). Note that instability in Fig. \ref{epsg0_01}(c,d) is present for some $u$ even when misfit is zero.
Responsible for this is the combined destabilizing effect of anisotropy and wetting stress, which together overweigh the stabilizing effect of the wetting layer; also see Sec. III.
Similar behaviour is observed for increasing $\mu$ while keeping thickness fixed.
We also notice that only $r=0.1$ case can be (probably) captured by longwave 
approximation, as $kh=2\pi ru=0.38\sim 1$ for $r=0.1,u=0.6$, and is even larger for other values of $r$ in Fig. \ref{epsg0_01}.
%\begin{figure}[H]
%%\centering
%\includegraphics[width=0.9\linewidth]{epsg0r01_1_3.eps}
%\caption{Neutral stability curves for $r=0.1$ (solid), $r=1$ (dash), $r=3$ (dot). $\epsilon_\gamma=0$. Domains of surface stability (instability) are
%marked by S (U). }
%\label{epsg0}
%\end{figure}
%
In order to characterize the horizontal spacing between two stability domains in Fig. \ref{epsg0_01}(c,d), in Fig. \ref{spacing} we plot the neutral stability curve corresponding to the level $\epsilon=0$.
It can be seen that for all reasonable $r$ this spacing does not exceed 0.3. For comparison, the case of slightly larger anisotropy is also shown. 

To summarize, we considered  all combinations of wetting interaction (through the exponential two-layer model), lattice-mismatch and wetting strains, and strong anisotropy.
Our results demonstrate complicated linear stability of ultrathin films ($h\sim 1:5$ wetting lengths). In particular, we show that extremely thin 
($h\sim 1:2$ wetting lengths), unstressed wetting films are not destabilized by arbitrarily strong anisotropy. Anisotropic, stressed wetting films
are destabilized by any level of mismatch stress, but only in the narrow range of perturbation wavenumbers. Such films can remain stable with respect to 
short-wavelength perturbations when they are very thin and at any reasonable mismatch stress level. Our final remark concerns two-dimensional surfaces and
corresponding surface energy anisotropies $\gamma_t(h_x,h_y)$ of the generic form \rf{1.2} (where contribution in the $y$-direction is additive, as is commonly assumed).
We conjecture that, if the surface orientation (of a thick film) is still one of the high-symmetry crystallographic orientations, such as [001] or [111], then 
the effect of such in-plane anisotropy is \emph{nonlinear} and thus the latter anisotropy will not affect the results. This can be qualitatively understood, for instance, 
by following the analysis leading to Eq. (15) in Ref. \cite{GolovinPRB2004} while accounting for the nonlinear nature of the mixed derivative term in Eq. \rf{1.4e} and the form of Eq. \rf{1.4fp}.
Due to complexity of formulation and derivation, the exact proof is beyond the scope of this note.
The results are also unchanged if the surface is two-dimensional but in-plane anisotropy is zero.

\acknowledgements
M.K. acknowledges the support of WKU Faculty Scholarship Council via grants 10-7016 and 10-7054, and thanks Brian J. Spencer (University at Buffalo) for comments on the first draft of the manuscript.

\end{document}